\documentclass[10pt, conference]{IEEEtran}

\usepackage[table,xcdraw]{xcolor}
\usepackage[hidelinks]{hyperref}
\usepackage[utf8]{inputenc}
\usepackage{color,soul}
\usepackage{mathtools}
\usepackage{amssymb}
\usepackage{algorithm,algpseudocode}
\usepackage[per-mode=symbol,range-phrase=~--~]{siunitx}
\usepackage{flexisym}
\usepackage[export]{adjustbox}
\usepackage{footnote}
\usepackage{makecell}
\usepackage{xurl}
\usepackage{comment}
\usepackage{caption}
\usepackage{subcaption}
\usepackage{graphicx}
\usepackage[inline,shortlabels]{enumitem}
\usepackage{multirow}
\usepackage{balance}
\newcommand{\ADAPT}{\mbox{ADApt}}

\newcommand{\CORE}{\mathtt{CORE}}
\newcommand{\CPU}{\mathtt{CPU}}
\newcommand{\MEM}{\mathtt{MEM}}

\newcommand{\sched}{\mathtt{sched}}

\newcommand{\REQ}{\mathtt{req}}
\newcommand{\RepSet}{\mathcal{R}}

\begin{document}

\title{\ADAPT{}: Edge Device Anomaly Detection\\and Microservice Replica Prediction}
\author{Narges Mehran\IEEEauthorrefmark{1}\IEEEauthorrefmark{2}, Nikolay Nikolov\IEEEauthorrefmark{3}, Radu Prodan\IEEEauthorrefmark{4}\IEEEauthorrefmark{5}, Dumitru Roman\IEEEauthorrefmark{3}, Dragi Kimovski\IEEEauthorrefmark{5},\\ Frank Pallas\IEEEauthorrefmark{2}, Peter Dorfinger\IEEEauthorrefmark{1}\\
\IEEEauthorblockA{\IEEEauthorrefmark{1} Intelligent Connectivity, Salzburg Research Forschungsgesellschaft mbH, Austria}
\IEEEauthorblockA{\IEEEauthorrefmark{2} Faculty of Digital and Analytical Sciences, Paris Lodron University of Salzburg, Austria}
\IEEEauthorblockA{\IEEEauthorrefmark{3} Sustainable Communication Technologies, SINTEF Digital, SINTEF AS, Oslo, Norway}
\IEEEauthorblockA{\IEEEauthorrefmark{4} Institute of Computer Science, University of Innsbruck, Austria}
\IEEEauthorblockA{\IEEEauthorrefmark{5} Institute of Information Technology, University of Klagenfurt, Austria}
}
\maketitle

\begin{abstract}
The increased usage of Internet of Things devices at the network edge and the proliferation of microservice-based applications create new orchestration challenges in Edge computing. These include detecting overutilized resources and scaling out overloaded microservices in response to surging requests. This work presents \emph{\ADAPT{}}, an extension of the ADA-PIPE tool developed in the DataCloud project, by monitoring Edge devices, detecting the utilization-based anomalies of processor or memory, investigating the scalability in microservices, and adapting the application executions. To reduce the overutilization bottleneck, we first explore monitored devices executing microservices over various time slots, detecting overutilization-based processing events, and scoring them. Thereafter, based on the memory requirements, \ADAPT{} predicts the processing requirements of the microservices and estimates the number of replicas running on the overutilized devices. The prediction results show that the gradient boosting regression-based replica prediction reduces the $\mathrm{MAE}$, $\mathrm{MAPE}$, and $\mathrm{RMSE}$ compared to others. Moreover, \ADAPT{} can estimate the number of replicas close to the actual data and reduce the $\CPU$ utilization of the device by \mbox{\qtyrange{14}{28}{\percent}}. 
\end{abstract}

\textcolor{red}{\scriptsize 2025 IEEE. Personal use of this material is permitted. Permission from IEEE must be obtained for reprinting/republishing/redistributing/reusing this work in other works.}
\section{Introduction} \label{sec:intro}
Effective management of microservice applications comes with unique orchestration challenges, such as utilization-based anomaly detection and scaling out overloaded microservices in response to increasing load~\cite{joseph2020intma}. The rapid proliferation of the Internet of Things (IoT) devices is anticipated to exceed \num{32} billion by \num{2030}\footnote{\url{https://statista.com/statistics/1183457/iot-connected-devices-worldwide}}, which fuels the expanded use of Edge devices for the execution of distributed microservices. Moreover, monitoring the utilization of virtual machines on the Cloud is more challenging than monitoring dedicated physical machines on the Edge due to the extensive use of virtualization~\cite{pourmajidi2017challenges}. Nevertheless, using Edge computing for microservices requires efficiently utilizing large numbers of heterogeneous, resource-constrained computing resources. Previous work~\cite{mehran2023comparison} explored microservice scaling on provisioned resources but did not consider Edge device anomalies or perform resource requirements prediction considering the application and infrastructure online monitoring data. Moreover, traditional anomaly detection~\cite{bu2024sst} or scaling methods~\cite{park2024graph,rzadca2020autopilot} focus on resource predictions and rarely explore the case of clustering the resources and detecting the overloaded devices before estimating requirements. Therefore, we investigate \emph{overutilization}-based \emph{Edge device anomaly detection} and the \emph{horizontal scaling} of containerized microservices on underutilized devices.

\emph{Example:} Table~\ref{tbl:example} shows three microservices running on two devices and their requirements of processing cores $\CORE$ and memory $\MEM$ (in \si{\giga\byte}). Figure~\ref{fig:example:anomal} presents a device's $d_0$ $\CORE$ and $\MEM$ resources experiencing varying utilization. Over the ten time slots, its processor is not anomalous, but memory presents overutilization anomalies (denoted as \num{1}) in \qty{80}{\percent} of the instances. We observe a direct correlation between the microservices' $\CORE$ and $\MEM$ requirements, motivating the need to explore a prediction model addressing their horizontal scaling~\cite{horn2022multi} to balance the varying request rates on the devices $\left\{d_0,d_1\right\}$. Thus, we replicate the microservice $m_0$ running on the anomalous device $d_0$ to satisfy peaks in user requests based on the ratio between the initial ($\CORE(m_0,t)=4$) and required computational requirements ($\CORE(m_0,t^{\prime})=3$): $m_0: \left\lceil{\frac{\SI{4}{}}{\SI{3}{}}}\right\rceil =\num{2};\ m_1: \left\lceil\frac{\SI{2}{}}{\SI{2}{}}\right\rceil=\num{1};\ m_2: \left\lceil\frac{\SI{3}{}}{\SI{3}{}}\right\rceil=\num{1}.$
\begin{table}[t]
\resizebox{.48\linewidth}{!}{
\begin{minipage}[]{.48\linewidth}
\caption{Exemplified microservices, devices, and resource requirements.}
\label{tbl:example}
\begin{tabular}{|@{}c@{}|@{}c@{}|@{}c@{}|@{}c@{}|}
 \hline
 \makecell{\emph{Micro-}\\\emph{service}} & \emph{Device} & \makecell{\emph{$\CORE\ \REQ$}\\(\si{\#})} & \makecell{\emph{$\MEM$ $\REQ$}\\(\si{\giga\byte})}\\\hline \hline
   $m_0$  & $d_0$ & \num{4} &  \num{4}\\\hline
   $m_1$  & $d_1$ & \num{2} &  \num{2}\\\hline
   $m_2$  & $d_0$ & \num{3} &  \num{3}\\\hline
\end{tabular}
\end{minipage}
}
\begin{minipage}[]{.48\linewidth}    \includegraphics[width=\linewidth]{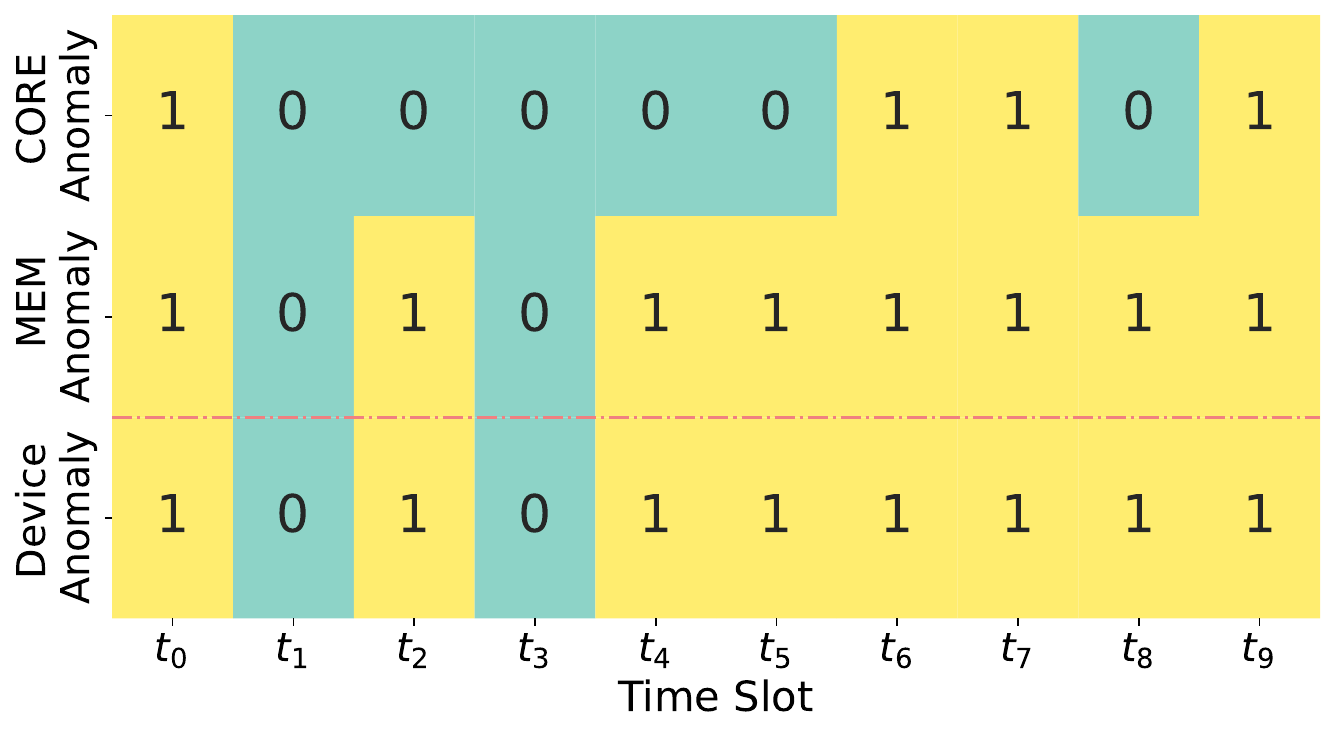}
    \captionof{figure}{Device $d_0$ overutilization anomaly in ten time slots.}
    \label{fig:example:anomal}
\end{minipage}
\end{table}

\emph{Method:} We address the device anomaly detection through $\CORE$ and $\MEM$ utilization of the Edge device. Thereafter, we investigate the scalability problem through \emph{$\CORE$ requirement predictions} employing a machine learning (ML) model involving two features:
\textit{1)} \emph{$\MEM$ requirement} defining the amount of $\MEM$ needed by each containerized microservice on the Edge machines; and \textit{2)} \emph{$\CORE$ requirement} defining the processing needed for each containerized microservice on the Edge machines. 
We apply ML models to predict $\CORE$ requirements based on the $\MEM$ and estimate the number of microservice replicas to support stochastic changes due to the dynamic user requirements. Recently, there has been a growing interest in the potential of using learning models for tabular data~\cite{gorishniy2021revisiting}. However, as explored in this work, tree-based ML models, such as gradient boosting regression (GBR), which is often used with tabular data, can outperform other learning methods~\cite{grinsztajn2022tree}. 

\emph{Contributions:} In this eight-section work, we present \ADAPT{} as an extension of the \mbox{ADA-PIPE} tool developed throughout the DataCloud project~\cite{roman2022big} with the following contributions: 
\begin{itemize}[leftmargin=*,align=left]
    \item Design of an Edge device anomaly detection model based on k-means clustering model for overutilization detection;
    \item Design of GBR, bagging regression (BR), and multilayer perceptron (MLP) models for predicting the number of microservice replicas running on overutilized devices;
    \item Empirical analysis of the \ADAPT{} in terms of its device utilization anomaly detection and accuracy improvement of the gradient boosting model for replica prediction;
    \item GBR-based \ADAPT{} reduces $\mathrm{MAE}$ = $\num{0.038}$, $\mathrm{MAPE}$ = $\num{0.002}$, and $\mathrm{RMSE}$ = $\num{0.196}$ compared to the BR and MLP.
\end{itemize}

\section{Related Work} \label{sec:related}
This section reviews the state-of-the-art infrastructure monitoring, anomaly detection, and autoscaling analysis.

\subsubsection*{Monitoring}
Prometheus extracts time-series data and leverages the PromQL query language to track metrics~\cite{turnbull2018monitoring}. Netdata provides the pre-built dashboards facilitating issue identification and data-driven decision-making~\cite{netdata1}. For Google Cloud users, Stackdriver offers a tightly integrated logging and monitoring method. cAdvisor~\cite{cadvisor1}, an open-source development by Google, collects and processes key metrics such as CPU, memory, storage, and network usage of containers. Prometheus has integrated support for cAdvisor~\cite{cadvisor2}, enabling users to configure Prometheus to scrape cAdvisor metrics. 

\subsubsection*{Anomaly detection}
Ahmad and Lavin~\cite{ahmad2017unsupervised} introduced the Numenta anomaly benchmark, designed to offer a controlled and reproducible environment with open-source tools for evaluating and measuring anomaly detection in streaming data. This method aims to adequately test and score real-time anomaly detectors' efficacy. Zhang et al.~\cite{zhang2018perfinsight} proposed a Mann-Kendall-based method that models entropy-based feature selection of transformed metrics. This method aims to improve the efficiency of model training and anomaly detection and reduce false positives in the detection phase. 

\subsubsection*{Autoscaling}
Park et al.~\cite{park2024graph} presented a graph neural network-based proactive resource autoscaling method for minimizing total processing resources while satisfying end-to-end latency.
Toka et al.~\cite{toka2021machine} presented a proactive scaling method, including multiple ML-based forecast models to optimize Edge resource over-provisioning and service level agreement.

\subsubsection*{Contribution} Related methods are designed as monitoring toolkits and resource analysis techniques or load prediction methods. \ADAPT{} extends these methods by exploring a monitoring toolkit to provide the required online metrics for Edge device anomaly detection and resource requirements prediction of microservices. Finally, \ADAPT{} method adapts the scheduling of microservices on the computing resources based on overutilization event detection using the k-means clustering model and replica prediction using the GBR, BR, and MLP machine learning regressions. 

\section{Model} \label{sec:model}
\subsection{Microservice, Edge device, and scheduling model}
\paragraph*{Bag of microservices} $\mathcal{B} =\left(\mathcal{M},\mathcal{S},\RepSet\right)$ with \emph{independent microservices} \mbox{$\mathcal{M} =\left\{m_i\ |\ 0 \leq i < \mathcal{N}_{\mathcal{M}}\right\}$} requiring a minimum number of cores $\CORE\left(m_i\right)$ and memory $\MEM\left(m_i\right)$ (in \si{\giga\byte}): $\REQ\left(m_i\right) = \left(\CORE\left(m_i\right), \MEM\left(m_i\right)\right)$, requested by users \mbox{$\mathcal{S} =\left\{s_i\ |\ 0 \leq i < \mathcal{N}_{\mathcal{S}}\right\}$}. Moreover, we represent microservice \emph{replicas} \mbox{$\RepSet=\{\RepSet_i| 0 \leq i < \mathcal{N}_{\mathcal{M}}\}$}, where $\RepSet_i$ corresponds to the required number of replicas for a microservice $m_i$.

\paragraph*{\emph{Edge devices}} $\mathcal{D} = \left\{d_j | 0 \leq j < \mathcal{N}_{\mathcal{D}}\right\}$  with $\CORE_j$ \emph{cores} and $\MEM_j$ \emph{memory} (in \si{\giga\byte}): \mbox{$d_j = \left(\CORE_j,\MEM_j\right)$}. 
\emph{Utilization} of a device $d_j$ at time $t$ is the $\CORE$ and $\MEM$ percentages (i.e., $\mathcal{U}\left(\CORE_j, t\right)$ and $\mathcal{U}\left(\MEM_j, t\right)$) used by microservice executions.

\paragraph*{\emph{Schedule}} $d_j$ = $\sched\left(m_i,t\right)$ is a mapping of a microservice replica $m_i$ to Edge device $d_j$ at time instance $t$. 

\subsection{Edge device anomaly detection model}
\subsubsection{Trace model} comprises Edge computing devices $d_j$ and resource \emph{utilization} based on the percentage of processing $\CORE$ or memory $\MEM$ at time $t$.

\subsubsection{K-means clustering model}
We define a model to classify the monitoring resources based on their utilization events. Among the widely used clustering algorithms, unsupervised learning-based methods such as k-means iterative clustering~\cite{lloyd1982least,elkan2003using} perform faster than hierarchical clustering such as tree-like structural one\footnote{\url{https://www.ibm.com/think/topics/k-means-clustering}}.
The k-means clustering maintains a set of $k$ centroids representing the clusters by utilizing the Elbow method and groups the monitoring utilization data (so-called data points) into clusters. 
\paragraph*{Elbow method} operates on the principle of conducting k-means clustering on a range of clusters (e.g., $k\in [1,10]$). Following each value in the range, we calculate the sum of squared distances from each monitoring data to its assigned centroid, known as distortions. By plotting these distortions to inspect the output curve, the elbow, resembling a bend of an arm or point of inflection, indicates the optimal $k$. 
\paragraph*{Clustering method} partitions the monitoring data of a resource into $k$ clusters, assigning each data point to the cluster with the closest mean (centroid), which serves as the representative of that cluster. Subsequently, the algorithm ranks the distances of each data point and identifies the $k$ nearest neighbors based on the shortest distances. If $k=2$, a monitoring data point is classified into the overutilization cluster if its proximity to the corresponding centroid is greater than that to the centroid of the under/full-utilization cluster. 

\subsection{Microservice replica prediction model}
\subsubsection{Trace model} comprises microservices $m_i\in\mathcal{M}$, its processing $\CORE\left(m_i\right)$ and memory $\MEM\left(m_i\right)$ requirements, and the number of microservice replicas $\RepSet_i$.

\subsubsection{ML models}
Boosting sequentially trains the models, with each model learning from the errors of its predecessor. In contrast, the bagging involves training multiple models independently and in parallel, each on a randomly selected subset of the data trace~\cite{ganaie2022ensemble}. Furthermore, bagging typically employs simple averaging to combine the model's predictions, whereas boosting assigns weights to models based on their predictive accuracy. In this work, an MLP model is a feedforward artificial neural network composed of an input, a hidden, and an output layer. The MLP applies initial weights to input data, calculates a weighted sum, and transforms it using an activation function akin to perceptron.

\section{Problem Definition} \label{sec:probdef}
This section defines the Edge device anomaly detection and microservice replica prediction problem. We assume that a microservice $m_i$ scheduled on an Edge device $d_j$ receives several incoming requests. Then, we explore the \emph{anomaly likelihood} $\mathcal{A}\left(d_j,t^{\prime}\right)$ of an Edge device and the \emph{replica set} $\RepSet_{i}$ of microservice $m_i$ at a future time instance $t^{\prime}$ based on time instance $t$. 
Therefore, one of the objectives of this paper is to devise an ML-based clustering model that predicts the anomaly likelihood $\mathcal{A}\left(d_j,t^{\prime}\right)$ of a device $d_j$ at a future time instance $t^{\prime}$, considering its scheduled microservices $\sched(m_i,t)$ and their replica set $\RepSet_i(t)$. 
More specifically, we investigate the anomaly score of the latest monitoring data compared to the historical resource usage. Moreover, if the anomaly score exceeds the availability thresholds $\theta_{\CORE}$ and $\theta_{\MEM}$, it is more likely that the monitoring data consists of an overutilization anomaly not found in the training data. 
Thereafter, the second objective of this paper is to explore an ML regression model that predicts the minimum number of replicas $\RepSet_i\left(t\right)$ that horizontally scales microservice $m_i$ to avoid anomalies. 

\paragraph*{Anomaly} \label{sssec:model_anomaly} 
$\mathcal{A}\left(d_j,t\right)$ of a device $d_j$ at time instance $t$ is the normalized score of raw anomaly $\mathcal{A}_{raw}\left(d_j,t\right)$ based on the minimum and maximum anomalies observed during model training: 
$\mathcal{A}\left(d_j,t\right)$=$\frac{\left|\mathcal{A}_{raw}\left(d_j,t\right)-\Call{Min}{\mathcal{A}}\right|}{\left|\Call{Max}{\mathcal{A}} -\Call{Min}{\mathcal{A}}\right|},$
where the $\Call{Min}{\mathcal{A}}$ and $\Call{Max}{\mathcal{A}}$
respectively represents the minimum and maximum raw anomaly scores. The raw anomaly $\mathcal{A}_{raw}\left(d_j,t\right)$ of a device $d_j$ at $t$ indicates its $\CORE$ or $\MEM$ utilization: $\mathcal{A}_{raw}\left(d_j,t\right)$ = $\mathcal{U}\left(\CORE_j, t\right) \ \lor\ \mathcal{U}\left(\MEM_j, t\right)$.

\paragraph*{Replica set} $\RepSet_{i}$ horizontally scales a microservice $m_i$ to $\RepSet_{i}\left(t^{\prime}\right)$ based on the multiplication of the current number of replicas $\RepSet_{i}\left(t\right)$ and the ratio between the current $\CORE\left(m_i,t\right)$ and required computational workload $\CORE\left(m_i,t^{\prime}\right)$:
$\RepSet_{i}(t^{\prime})$=$\left\lceil \RepSet_{i}(t)\cdot\frac{\CORE\left(m_i,t\right)}{\CORE\left(m_i,t^{\prime}\right)}\right\rceil$,
where the prediction model forecasts $\CORE\left(m_i,t^{\prime}\right)$ based on nearly correlated  $\MEM\left(m_i,t\right)$, to estimate $\RepSet_{i}\left(t^{\prime}\right)$. 

\section{Architecture Design} \label{sec:arch}
\begin{figure}[t]
    \centering
    \includegraphics[width=.75\columnwidth]{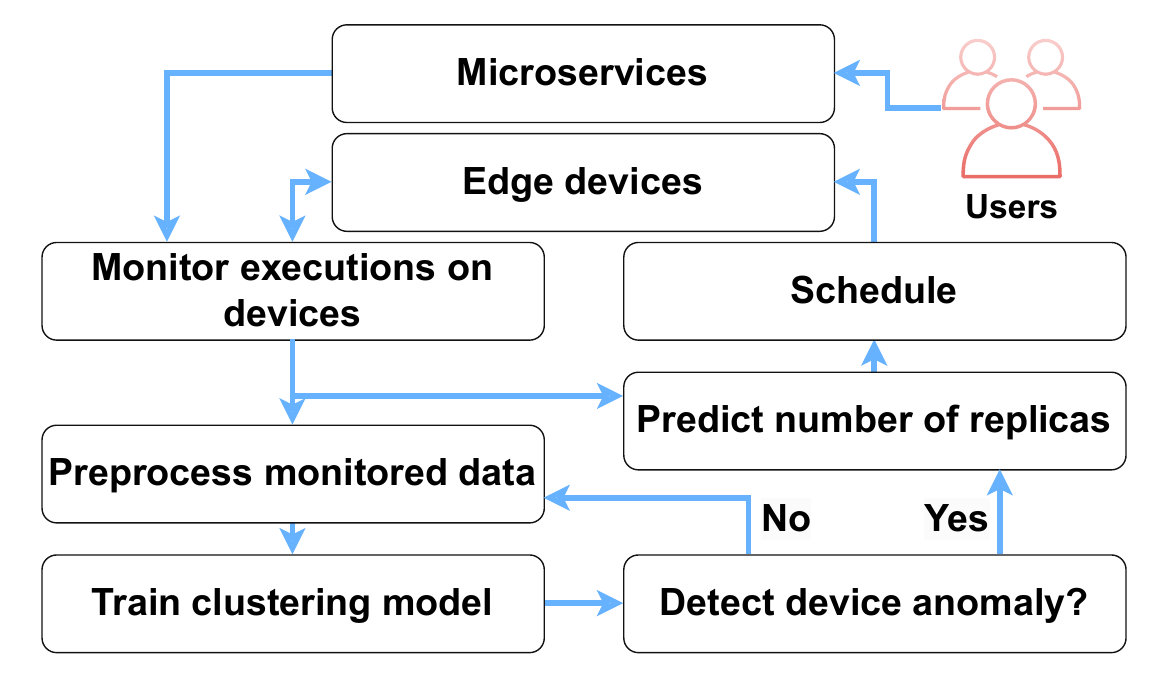}
    \caption{\ADAPT{} architecture.}
    \label{fig:arch}
\end{figure}
\ADAPT{} is an extension of the ADA-PIPE tool developed throughout the DataCloud project~\cite{roman2022big}. Figure~\ref{fig:arch} illustrates the \ADAPT{} components: \textit{1)} monitoring the microservice executions on the Edge devices, \textit{2)} data preprocessing, \mbox{\textit{3)} [re-]training} a clustering model on the preprocessed data, \textit{4)} detecting anomalous device utilization event, and \textit{5)} predicting the number of required replicas, and \textit{6)} adapting the initially scheduled devices. In detail, to record the service executions on the computing devices, the \texttt{Prometheus} monitoring system imports the \texttt{NetData} metrics~\cite{toka2021predicting}, such as processor and memory utilization, alongside the runtime of microservices~\cite{cadvisor1}. Afterward, \ADAPT{} preprocesses the monitoring data for the anomaly detection phase. The preprocessing phase creates a differenced, smoothed, and lagged data-trace.  
Afterward, \ADAPT{} trains a k-means model as a popular unsupervised clustering algorithm on the monitoring data. Furthermore, during every time interval, \ADAPT{} retrains the model based on the monitoring information (i.e., $\CORE$ and $\MEM$ utilization). Moreover, if a device's resource exceeds its utilization threshold, the \ADAPT{} re-schedules the service~\cite{mehran2022matching}.

\section{Experimental Design}\label{sec:experimntdesign} 
This section presents our experimental design for a testbed monitored during the runtime of microservices, the dataset preparation, and the tuning of hyperparameters. We monitored and implemented ML-based clustering and regression algorithms in \texttt{Python 3.9.13} on two devices with \num{10}-core Intel$^\circledR$ Core$^{(TM)}$ \mbox{i5-1335U} processor and \qty{16}{\giga\byte} of memory.

\subsection{Data preprocessing}
For the preprocessing stage, we sorted the monitoring data based on the task's earliest and finishing times by using the \texttt{Pandas.DataFrame}\footnote{\url{https://pandas.pydata.org/docs/reference/frame.html}} on our two-dimensional data structure. Moreover, we achieved differenced, smoothed, and lagged monitoring data by \texttt{diff}
, \texttt{rolling}
, \texttt{reindex}
, and \texttt{concat} methods 
 of \texttt{Pandas.DataFrame}.

\subsection{K-means model design}
We used the \texttt{KMeans} library from \texttt{sklearn} API to cluster the utilization of the device during the resource monitoring in a specific time interval\footnote{\url{https://github.com/DataCloud-project/ADA-PIPE/tree/main/detect-anomalies}}. 
We set $t_{trn}$ to \numrange{500}{3600}, showing how often to re-train the k-means model, and $\mathcal{N}_{trn}$ to the range of \qtyrange{500}{3600}{\hour}, showing that it trains the model on the last hour of data during each training interval. In this work, the \texttt{Netdata} collects runtime data from the range of \qtyrange{30}{60}{\day} of the utilized computing devices. \texttt{Netdata} chart contains $\CORE$ and $\MEM$ usage of a \texttt{Linux} system running on Edge device, and more specifically, its non-kernel user mode or dimension. $\CORE$ trace data is available through a public access point\footnote{\url{https://zenodo.org/records/14961415}\label{footnot:data}}. Moreover, we set the number of time instances \num{1}, \num{3}, and \num{5}, respectively, to difference, smooth, and lag the time series of monitoring data in the preprocessing stage.

\subsubsection{Elbow method}
\begin{figure}[t]
    \centering
    \subfloat[Elbow method estimating number of clusters.]{\includegraphics[width=0.4\columnwidth]{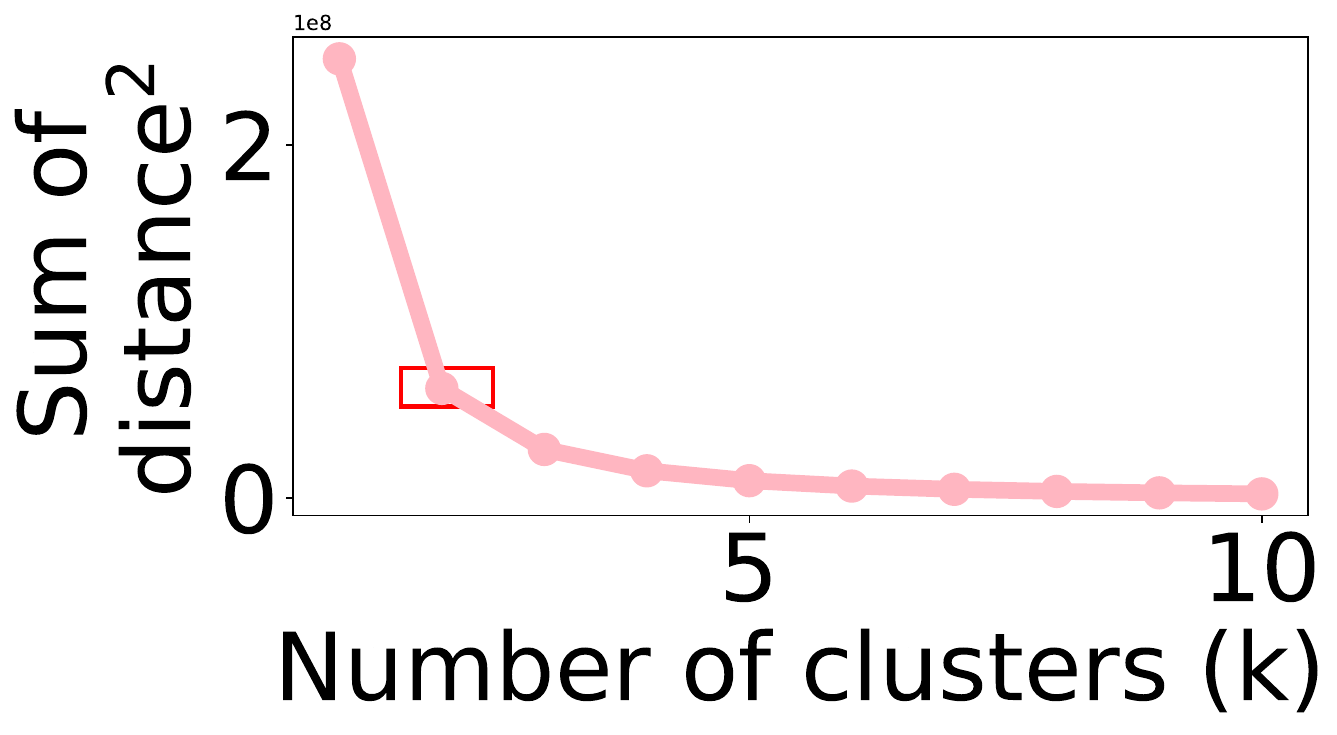}
    \label{fig:elbow}}
    \hspace{0.1cm}
    \subfloat[Two clusters based on $\mathcal{U}\left(\CORE_j, t\right)$.]{\includegraphics[width=0.4\columnwidth]{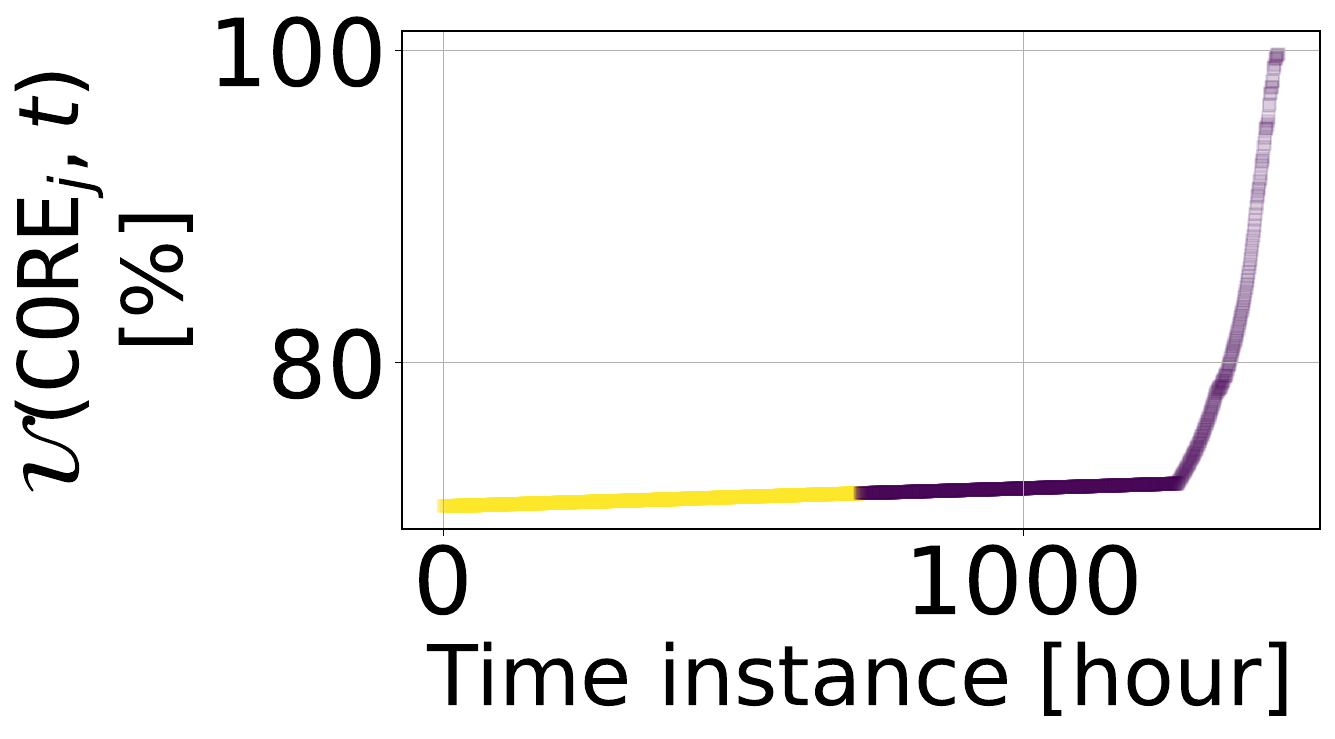}\label{fig:clustering}}
    \caption{Over- and full/under-utilization clusters in k-means-based anomaly detection model.}
\end{figure}
We utilized the Elbow method~\cite{schubert2023stop} to investigate the number of clusters within the range of \mbox{\numrange{1}{10}} to fit a proper learning model to the Edge device data-trace. Figure~\ref{fig:elbow} shows an elbow at $k=2$  (shown with a red rectangle) that defines the number of clusters to fine-tune the model's parameters.  
This comes from analyzing the sum of squared distances between data points and their cluster centroids. \ADAPT{} calculates the sum of squared distances by taking the squared differences between the corresponding coordinates of each data point and its assigned centroid. 

\subsubsection{Anomaly detection}
\ADAPT{} calculates the device anomaly scores using an Euclidean distance metric through 
the \texttt{cdist} function from the \texttt{scipy.spatial.distance} library. 
As Figure~\ref{fig:clustering} shows, the k-means-based \ADAPT{} clusters of the $\CORE$ utilization of Edge device over a \SI{1440}{\hour} time interval with the cluster centroid coordinates $\left(\SI{1078.5}{\hour},\SI{74.43}{\percent}\right)$ of the overutilization cluster shown with purple color, and $\left(\SI{358.5}{\hour},\SI{71.21}{\percent}\right)$ of the full-utilization shown with yellow color. 
In this work, we set 
a threshold \SI{73}{\percent} of a device's maximum capacity (for number of $\CORE$ or $\MEM$ in \si{\giga\byte}). As the result of clustering shows, the Edge device is mostly in the overutilization event, which marks this device as an anomaly that requires devising a replication for its hosted microservices.


\subsection{ML hyperparameter design}
This section presents the learning procedure of fine-tuning and optimization of the hyperparameters of the GBR, BR, and MLP models based on three steps: exhaustive search, hyperparameter tuning, and hyperparameter configuration using \texttt{scikit-learn 1.5.1}~\cite{scikitlearn_api}. The script to run these models is available in GitHub public repository\footnote{\url{https://github.com/DataCloud-project/ADA-PIPE/tree/main/replica-prediction}}. We utilized two queries related to containerized microservice $\CPU$ and $\MEM$ average usage to collect the required \texttt{Prometheus} monitoring data for two months corresponding to \SI{1440}{\hour}\textsuperscript{\ref{footnot:data}}.
\begin{figure}[t]
    \centering
    \subfloat[GBR Loss.]{\includegraphics[width=.31\columnwidth]{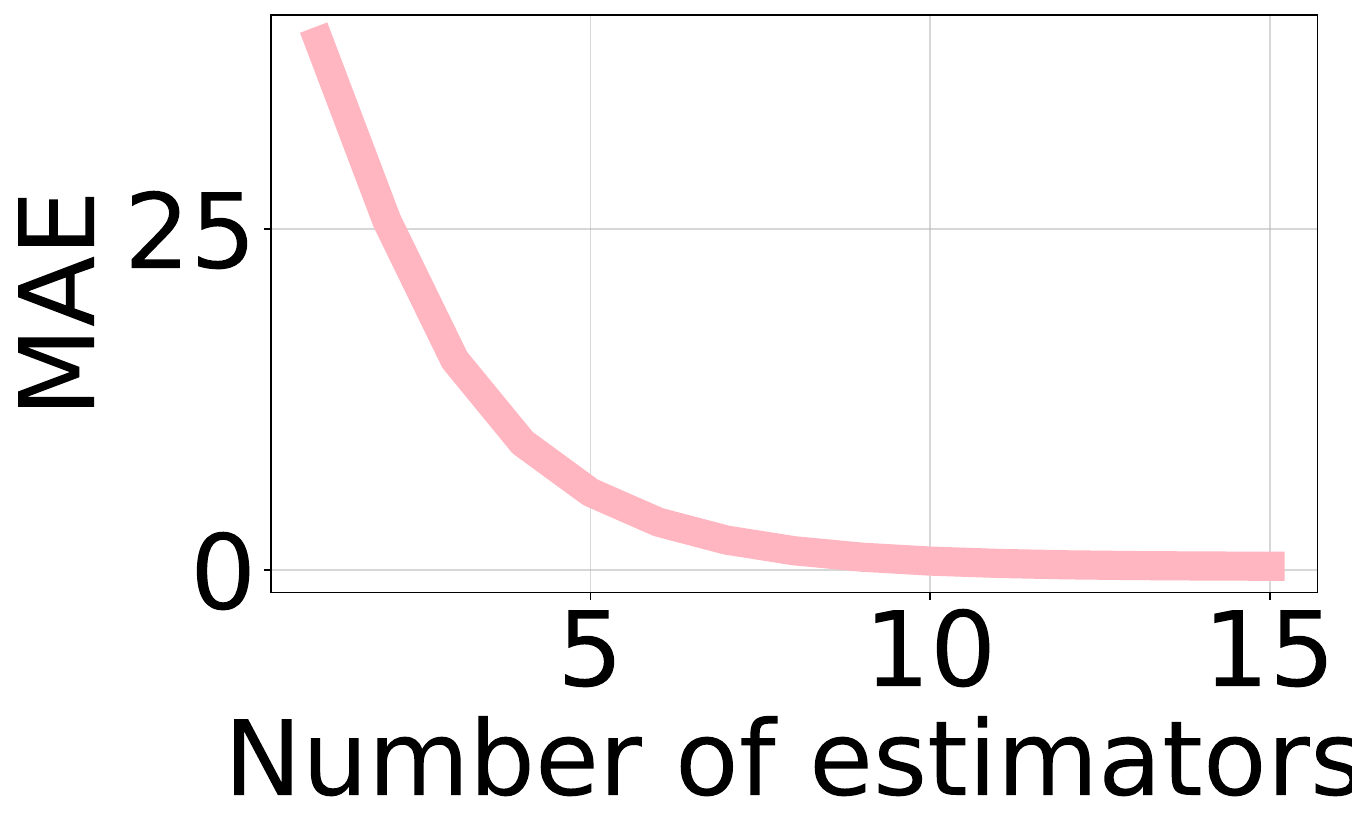}\label{fig:gbr-val}}\vspace{0.01cm}
    \subfloat[BR Loss.]{\includegraphics[width=.31\columnwidth]{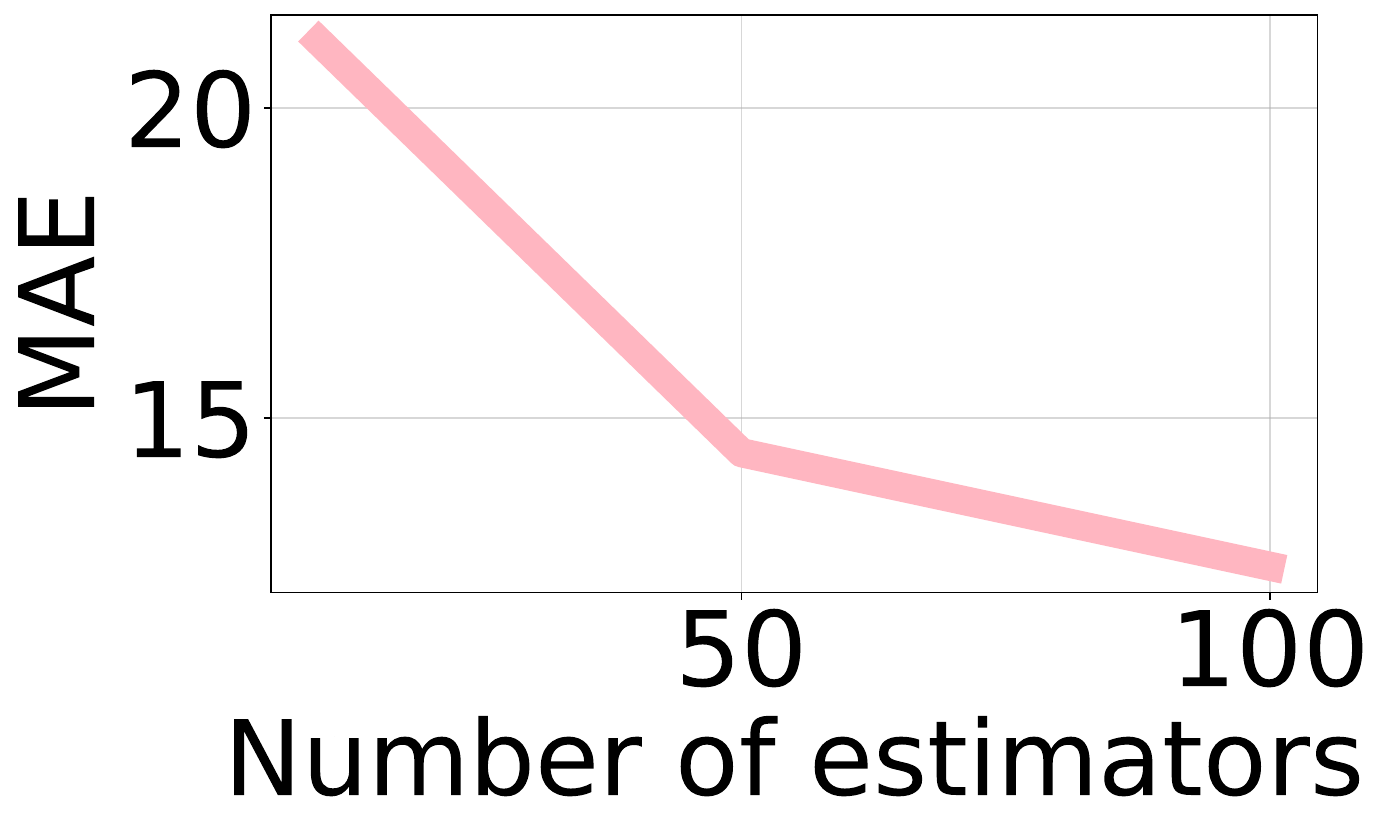}\label{fig:br-val}}\vspace{0.01cm}
    \subfloat[MLP Loss.]{\includegraphics[width=.31\columnwidth]{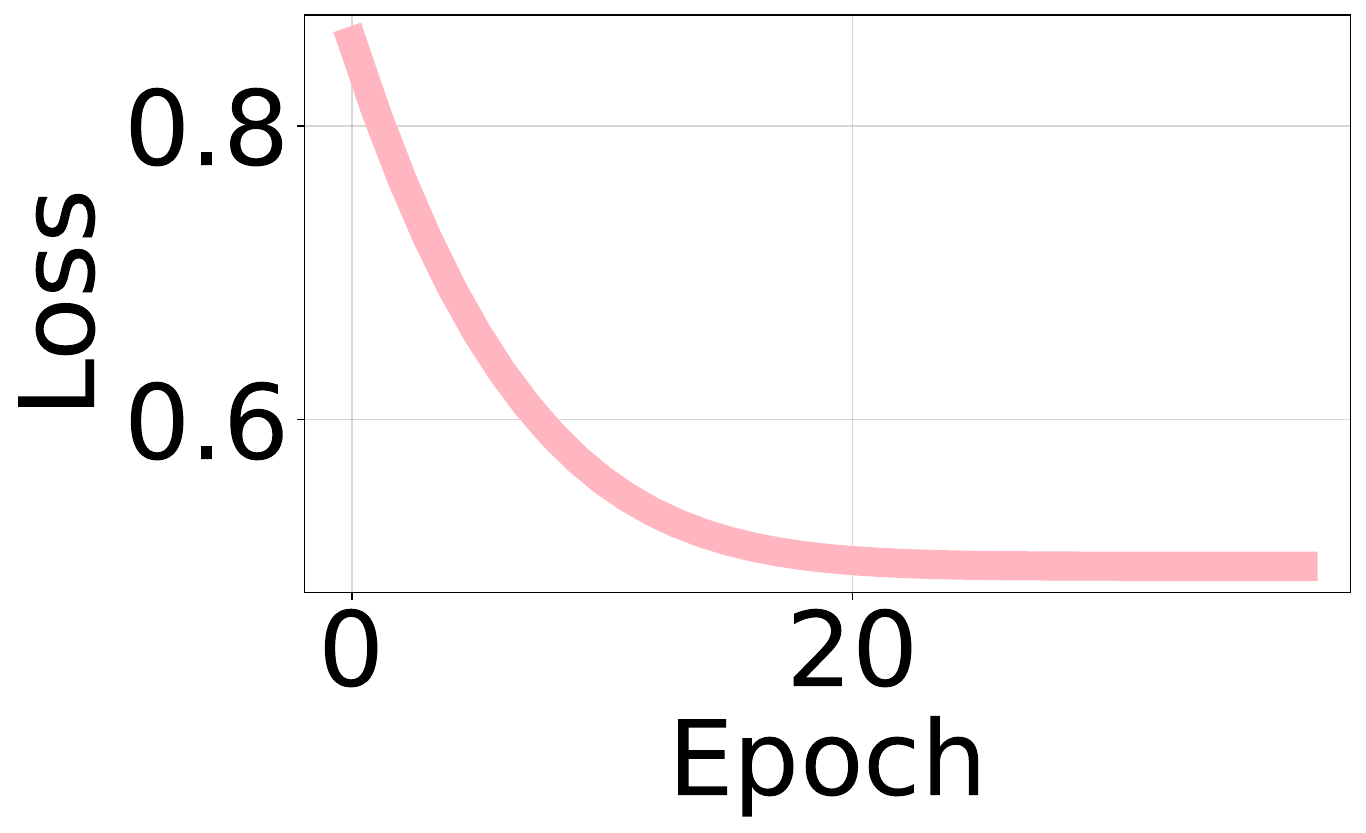}\label{fig:mlp-val}}
    \caption{Loss during the training iterations.}
\end{figure}
\subsubsection{Gradient boosting regressor} uses a learning curve to evaluate changes in training loss across iterations and considers the number of estimators and learning rate. An exhaustive search, implemented via \texttt{GridSearchCV} with \num{300} estimators and a learning rate of \num{0.02}, leads to overfitting. Therefore, we applied hyperparameter tuning to adjust the number of estimators (\numrange{10}{170}) and learning rate (\numrange{0.02}{0.4}), achieving convergence with faster training. The optimal configuration sets \num{15} estimators and a learning rate of \num{0.4}, improving the training score while avoiding overfitting and reducing training time. Figure~\ref{fig:gbr-val} illustrates the GBR iterative error correction procedure while reaching the maximum number of estimators.

\subsubsection{Bagging regressor} concurrently fits a set of models on random subsets of the data and averages the predictions of each sub-model\footnote{\url{https://scikit-learn.org/stable/modules/generated/sklearn.ensemble.BaggingRegressor.html}}. We optimized the model by using the \texttt{GridSearchCV} to modify the estimators in the range of \numrange{10}{100} and the maximum number of samples in \numrange{0.5}{5} for training each base estimator (see Figure~\ref{fig:br-val}). 

\subsubsection{Multilayer perceptron regressor} fits and optimizes a one-hidden-layer model by reducing the loss during the training iterations within the range of \numrange{1}{25} epochs via \texttt{HalvingRandomSearchCV} library (see Figure~\ref{fig:mlp-val}).

\subsection{Evaluation metrics}
We evaluate the \ADAPT{} ML-based models by five metrics.

\paragraph{Mean absolute error} ($\mathrm{MAE}$) represents the average sum of absolute differences between the predicted number of replicas $\RepSet_{i}\left(t^{\prime}\right)$ and $\RepSet_{i}\left(t\right)$, respectively, in testing and training. 
\paragraph{Mean absolute percentage error} ($\mathrm{MAPE}$) quantifies the prediction accuracy of an ML model. 
\paragraph{Root mean squared error} ($\mathrm{RMSE}$) quantifies the prediction accuracy of an ML model based on the square root of the mean squared error ($\mathrm{MSE}$).

\paragraph{Number of replicas} 
$\RepSet_{i}\left(t^{\prime}\right)$ defined in Section~\ref{sec:model}.

\paragraph{Utilization of device} $\mathcal{U}\left(\CORE_j, t^{\prime}\right)$ and $\mathcal{U}\left(\MEM_j,t^{\prime}\right)$  defined in Section~\ref{sec:model}.

\section{Experimental Results} \label{sec:experimntresult}
\begin{table}[t]
\centering
\caption{\ADAPT{} prediction errors and training time.}
\label{tbl:errors}
\resizebox{0.8\columnwidth}{!}{
\begin{tabular}{|c||c|c|c|c|c|}
    \hline
    \makecell{\textit{Prediction model}} &
    $\mathrm{MAE}$ & $\mathrm{MAPE}$ & $\mathrm{RMSE}$&\makecell{\textit{Training time} [\si{\second}]}\\\hline\hline
    \textit{GBR}&\num{0.038}&\num{0.002}&\num{0.196}&\SI{0.2}{}\\   \hline
    \textit{BR}&\num{0.962}&\num{0.076}&\num{4.330}&\SI{0.3}{}\\   \hline
    \textit{MLP}&\num{1.123}&\num{0.177}&\num{4.611}&\SI{0.4}{}\\   \hline
\end{tabular}}
\end{table}
\subsubsection*{Number of replicas}  
Table~\ref{tbl:errors} shows that the GBR-based \ADAPT{} reduces $\mathrm{MAE}=\num{0.038}$, $\mathrm{MAPE}=\num{0.002}$, and $\mathrm{RMSE}=\num{0.196}$ compared to the BR- and MLP-based \ADAPT{}. Moreover, the BR model outperforms the MLP for training time and errors. 
\begin{table}[t]
\centering
\caption{\centering Comparison of \ADAPT{} and \textit{without-prediction}.}
\label{tbl:}
\subfloat[Number of replicas $\RepSet_i\left(t\right)$ and $\RepSet_i\left(t^{\prime}\right)$.]{
    \label{tbl:compare_replica}
    \begin{tabular}{|@{}c@{}||@{}c@{}|@{}c@{}|}
    \hline
    \textit{Model} & \makecell{\textit{Without} \textit{prediction}} & \ADAPT{}\\
    \hline
    \hline
    \textit{GBR}&\num{42}&\num{43}\\
    \hline
    \textit{BR}&\num{42}&\num{41}\\
    \hline
    \textit{MLP}&\num{42}&\num{38}\\
    \hline
    \end{tabular}
}
\quad
\subfloat[Utilization of Edge device.]{
    \label{tbl:utilization}
    \begin{tabular}{|@{}c@{}||@{}c@{}|@{}c@{}|}
    \hline
    {\textit{Resource}} & \makecell{\textit{Without}\\
     \textit{prediction}
    } & \makecell{\ADAPT{}}\\
    \hline
    \hline
    $\CORE_j$&\numrange{70}{98}&\numrange{60}{70}\\
    \hline
    $\MEM_j$&\numrange{50}{70}&\numrange{45}{65}\\
    \hline
    \end{tabular}
}
\end{table}
Table~\ref{tbl:compare_replica} shows that the \ADAPT{} model estimates the number of replicas by following an almost close prediction to the without-prediction model. 

\subsubsection*{Utilization of a device}
Table~\ref{tbl:utilization} shows that \ADAPT{} reduces the utilization of a device by \mbox{\qtyrange{14}{28}{\percent}} and \mbox{\qtyrange{7}{10}{\percent}}. Moreover, the results show that applying device anomaly detection and replica prediction lessens the overutilization of computing resources through load balancing. 

\section{Conclusion and Future Work} \label{sec:conclusion}
We presented an ML-based clustering and boosting method to improve resource provisioning affected by stochastic changes due to users' requirements. Using the monitoring data, we investigated the performance evaluation of clustering, ensemble, and neural network models. We applied various ML models: \textit{1)} unsupervised k-means clustering to detect the Edge device overutilization anomaly, and \textit{2)} gradient boosting, bagging, and multilayer perceptron regressions to predict $\CORE$ based on $\MEM$ requirements for scaling the microservices on the overloaded Edge. 
The experimental results show that \ADAPT{} can detect the overutilized Edge devices based on $\CORE$ utilization. Moreover, the results show that the GBR-based replica prediction reduces the $\mathrm{MAE}$, $\mathrm{MAPE}$, and $\mathrm{RMSE}$ compared to BR and MLP models. Besides, \ADAPT{} estimates the number of replicas for each microservice close to the actual values without any prediction. In the future, we plan to integrate other ML models with green Edge monitoring data while preserving data privacy~\cite{pallas2020fog}. 

\section*{Acknowledgements} This work received funding from the Land Salzburg 20204-WISS/263/6-6022 (EXDIGIT), Horizon Europe projects 101093202 (Graph-Massivizer), 101189771 (DataPACT), 101070284 (enRichMyData), 101093216 (UPCAST), 101135576 (INTEND), and Austrian Research Promotion Agency 909989 (AIM AT Stiftungsprofessur für Edge AI).

\balance
\bibliography{ref}

\begin{thebibliography}{10}

\bibitem{joseph2020intma}
Christina~Terese Joseph and K~Chandrasekaran.
\newblock {IntMA}: Dynamic interaction-aware resource allocation for containerized microservices in cloud environments.
\newblock {\em Journal of Systems Architecture}, 111:101785, 2020.

\bibitem{pourmajidi2017challenges}
William Pourmajidi, John Steinbacher, Tony Erwin, and Andriy Miranskyy.
\newblock On challenges of cloud monitoring.
\newblock In {\em Proceedings of the 27th Annual International Conference on Computer Science and Software Engineering}, pages 259--265, 2017.

\bibitem{mehran2023comparison}
Narges Mehran, Arman Haghighi, Pedram Aminharati, Nikolay Nikolov, Ahmet Soylu, Dumitru Roman, and Radu Prodan.
\newblock Comparison of microservice call rate predictions for replication in the cloud.
\newblock In {\em Proceedings of the IEEE/ACM 16th International Conference on Utility and Cloud Computing}, pages 1--7, 2023.

\bibitem{bu2024sst}
Shilei Bu, Minpeng Jin, Jie Wang, Yulai Xie, and Liangkang Zhang.
\newblock {SST-LOF}: Container anomaly detection method based on singular spectrum transformation and local outlier factor.
\newblock {\em IEEE Transactions on Cloud Computing}, 2024.

\bibitem{park2024graph}
Jinwoo Park, Byungkwon Choi, Chunghan Lee, and Dongsu Han.
\newblock Graph neural network-based {SLO}-aware proactive resource autoscaling framework for microservices.
\newblock {\em IEEE/ACM Transactions on Networking}, 32(4):3331--3346, 2024.

\bibitem{rzadca2020autopilot}
Krzysztof Rzadca, Pawel Findeisen, Jacek Swiderski, Przemyslaw Zych, Przemyslaw Broniek, Jarek Kusmierek, Pawel Nowak, Beata Strack, Piotr Witusowski, Steven Hand, et~al.
\newblock Autopilot: workload autoscaling at google.
\newblock In {\em Proceedings of the Fifteenth European Conference on Computer Systems}, pages 1--16, 2020.

\bibitem{horn2022multi}
Angelina Horn, Hamid~Mohammadi Fard, and Felix Wolf.
\newblock Multi-objective hybrid autoscaling of microservices in kubernetes clusters.
\newblock In {\em Euro-Par 2022: Parallel Processing: 28th International Conference on Parallel and Distributed Computing}, pages 233--250. Springer, 2022.

\bibitem{gorishniy2021revisiting}
Yury Gorishniy, Ivan Rubachev, Valentin Khrulkov, and Artem Babenko.
\newblock Revisiting deep learning models for tabular data.
\newblock {\em Advances in Neural Information Processing Systems}, 34:18932--18943, 2021.

\bibitem{grinsztajn2022tree}
L{\'e}o Grinsztajn, Edouard Oyallon, and Ga{\"e}l Varoquaux.
\newblock Why do tree-based models still outperform deep learning on typical tabular data?
\newblock {\em Advances in Neural Information Processing Systems}, 35:507--520, 2022.

\bibitem{roman2022big}
Dumitru Roman, Radu Prodan, Nikolay Nikolov, Ahmet Soylu, Mihhail Matskin, Andrea Marrella, Dragi Kimovski, Brian Elvesæter, Anthony Simonet-Boulogne, Giannis Ledakis, Hui Song, Francesco Leotta, and Evgeny Kharlamov.
\newblock Big data pipelines on the computing continuum: Tapping the dark data.
\newblock {\em Computer}, 55(11):74--84, 2022.

\bibitem{turnbull2018monitoring}
James Turnbull.
\newblock {\em Monitoring with Prometheus}.
\newblock Turnbull Press, 2018.

\bibitem{netdata1}
{Netdata, Inc.}
\newblock {Using Netdata with Prometheus}.
\newblock \url{https://learn.netdata.cloud/docs/exporting-data/prometheus}, 2025.
\newblock [Online; accessed Mar. 2025].

\bibitem{cadvisor1}
{Google team}.
\newblock {Analyzes resource usage and performance characteristics of running containers}.
\newblock \url{https://github.com/google/cadvisor}, 2025.
\newblock [Online; accessed Mar. 2025].

\bibitem{cadvisor2}
{Prometheus Authors}.
\newblock {Monitoring docker container metrics using cadvisor}.
\newblock \url{https://prometheus.io/docs/guides/cadvisor/}, 2025.
\newblock [Online; accessed Mar. 2025].

\bibitem{ahmad2017unsupervised}
Subutai Ahmad, Alexander Lavin, Scott Purdy, and Zuha Agha.
\newblock Unsupervised real-time anomaly detection for streaming data.
\newblock {\em Neurocomputing}, 262:134--147, 2017.

\bibitem{zhang2018perfinsight}
Xiao Zhang, Fanjing Meng, and Jingmin Xu.
\newblock Perfinsight: A robust clustering-based abnormal behavior detection system for large-scale cloud.
\newblock In {\em 2018 IEEE 11th International Conference on Cloud Computing (CLOUD)}, pages 896--899. IEEE, 2018.

\bibitem{toka2021machine}
L{\'a}szl{\'o} Toka, Gergely Dobreff, Bal{\'a}zs Fodor, and Bal{\'a}zs Sonkoly.
\newblock Machine learning-based scaling management for kubernetes edge clusters.
\newblock {\em IEEE Transactions on Network and Service Management}, 18(1):958--972, 2021.

\bibitem{lloyd1982least}
Stuart Lloyd.
\newblock Least squares quantization in {PCM}.
\newblock {\em IEEE transactions on information theory}, 28(2):129--137, 1982.

\bibitem{elkan2003using}
Charles Elkan.
\newblock Using the triangle inequality to accelerate k-means.
\newblock In {\em Proceedings of the 20th international conference on Machine Learning (ICML-03)}, pages 147--153, 2003.

\bibitem{ganaie2022ensemble}
Mudasir~A Ganaie, Minghui Hu, Ashwani~Kumar Malik, Muhammad Tanveer, and Ponnuthurai~N Suganthan.
\newblock {Ensemble deep learning: A review}.
\newblock {\em Engineering Applications of Artificial Intelligence}, 115:105151, 2022.

\bibitem{toka2021predicting}
Laszlo Toka, Gergely Dobreff, David Haja, and Mark Szalay.
\newblock Predicting cloud-native application failures based on monitoring data of cloud infrastructure.
\newblock In {\em 2021 IFIP/IEEE International Symposium on Integrated Network Management (IM)}, pages 842--847. IEEE, 2021.

\bibitem{mehran2022matching}
Narges Mehran, Zahra~Najafabadi Samani, Dragi Kimovski, and Radu Prodan.
\newblock Matching-based scheduling of asynchronous data processing workflows on the computing continuum.
\newblock In {\em 2022 IEEE International Conference on Cluster Computing (CLUSTER)}, pages 58--70, 2022.

\bibitem{schubert2023stop}
Erich Schubert.
\newblock Stop using the elbow criterion for k-means and how to choose the number of clusters instead.
\newblock {\em ACM SIGKDD Explorations Newsletter}, 25(1):36--42, 2023.

\bibitem{scikitlearn_api}
{{Lars Buitinck, et al.}}
\newblock {API} design for machine learning software: experiences from the scikit-learn project.
\newblock In {\em ECML PKDD Workshop: Languages for Data Mining and Machine Learning}, pages 108--122, 2013.

\bibitem{pallas2020fog}
Frank Pallas, Philip Raschke, and David Bermbach.
\newblock Fog computing as privacy enabler.
\newblock {\em IEEE Internet Computing}, 24(4):15--21, 2020.

\end{thebibliography}
\bibliographystyle{unsrt}

\end{document}